\documentclass[aps,prd,twocolumn,floatfix]{revtex4}

\usepackage{graphicx}

\begin{document}








\newcommand{\be}{\begin{equation}}
\newcommand{\ee}{\end{equation}}
\newcommand{\ba}{\begin{eqnarray}}
\newcommand{\ea}{\end{eqnarray}}
\newcommand{\NL}{\nonumber \\}

\newcommand{\binom}[2]{ \left( \begin{array}{c} #1 \\ #2 \end{array} \right)}
\newcommand{\E}[1]{ \langle #1 \rangle }
\newcommand{\ov}[1]{ \overline{#1} }


%
\title{Chaotic Quantization of Four-Dimensional U(1) Lattice Gauge Theory}

\author{Tam\'as S.\ Bir\'o}
\affiliation{KFKI Res.\ Inst.\ Part.\ Nucl.\ Phys., 
             H-1525 Budapest 49, Hungary}

\author{Berndt M\"uller}
\affiliation{Physics Department, Duke University, Durham, NC 27708, USA}

\begin{abstract} 
We demonstrate that the quantized U(1) lattice gauge theory in four
Euclidean dimensions can be obtained as the long time average of the 
corresponding classical U(1) gauge theory in 4+1 dimensions. The 
Planck constant $\hbar$ is related to the excitation energy and 
the lattice constant of this classical template.
\end{abstract}

\maketitle

\section{Introduction}

The nonlinear classical dynamics of gauge fields is known to be 
strongly chaotic \cite{MST81,BMM95}. The classical SU(2) gauge
theory defined on a three-dimensional lattice has been studied
numerically in considerable detail and was shown to be a globally 
hyperbolic (Anosov) system \cite{chaos}. For such systems, any 
generic initial gauge field configuration wanders ergodically in 
the phase space of field configurations. This motion
leads in the infrared limit to a stationary 
distribution of lower dimensional configurations \cite{BMM01}. 
These, as a result of the higher dimensional {\em classical} 
dynamics, are distributed exactly as required by the vacuum state
of the Euclidean {\em quantum} theory on the lower dimensional 
space. 

This phenomenon, which has been called {\it chaotic quantization}
\cite{Beck95,BMM01}, can be simply considered as a physical 
realization of the method of stochastic quantization \cite{PW81}. 
It is a consequence of two deep relationships: (1) that the long 
time properties of certain deterministic classical systems can be 
described by the methods of statistical physics \cite{Dorfman}, and 
(2) the correspondence between classical statistics and Euclidean 
quantum mechanics. In the appendix we demonstrate this correspondence
for the quantum harmonic oscillator in detail.

According to the chaotic quantization scenario it is necessary for
the higher dimensional (second time) dynamics to be chaotic in order to
appear as a quantum theory in the lower (one-time) dimensions.
From the investigation of the U(1) gauge theory presented in this article
we shall see that the existence of the lattice is essential 
for this scenario to work in those cases when the dynamics would not
be chaotic in the weak coupling limit. The chaotic quantization may fail
for systems not being chaotic on finite lattices as well as in case
of observing too short second-time of the dynamical evolution.
This opens up towards the possibility that not all interactions,
in particular gravity, have to be quantized in nature.

This way the chaotic quantization of
gauge theories also lends support to 't Hooft's proposal \cite{tHooft} 
that quantum mechanics could arise from an underlying dissipative, 
yet microscopically deterministic classical dynamics. We will not
explore this speculative avenue here and confine ourselves to 
reporting our numerical evidence for a correspondence between
the classical U(1) gauge theory in 4+1 dimensions and the quantum
U(1) gauge theory in 4 dimensions.

The classical compact U(1) gauge theory without matter fields
is the simplest gauge theory still exhibiting chaotic dynamics in 
calculations on three-dimensional lattices \cite{U1}. Although its 
continuum limit is not chaotic, in contrast to non-Abelian gauge 
theories where chaos survives even in this limit, the classical 
dynamics of the compact U(1) lattice gauge field is chaotic at 
finite lattice spacing $a$ and sufficiently strong coupling $g$
($ga^2$ finite)\cite{BMM95}. As in the case of non-abelian gauge
theories few mode models, like the  $xy$-model \cite{XY-MODEL}
can be studied, a 3-link section of U(1) lattice gauge theory reveals
a system of harmonically coupled pendulums, easy to see numerically
drifting in chaotic behavior at high energy.
The dimensionless parameter controlling the strength
of chaoticity of the lattice gauge theory is $g^2 a E_p$, where 
$E_p$ is the average total energy per elementary lattice plaquette.

Our conjecture of the correspondence between the (4+1)-dimensional 
classical gauge theory with chaotic dynamics and the 4-dimensional 
Euclidean quantum field theory can be expressed in the following 
relation between the 4-dimensional Planck constant $\hbar$ and two 
physical parameters of the higher dimensional classical theory: 
the temperature $T$ and the lattice spacing \cite{BMM01}:
\be
 \hbar = a T. 
\ee
Our philosophy here views $\hbar$ as a constant of nature factorized 
into two underlying properties of the true (4+1)-dimensional world. 
An analogous case is the relation of classical electrodynamics 
factorizing the speed of light $c$ into electric and magnetic 
properties of the vacuum:
\be
1/c^2 = \epsilon_0 \mu_0.
\ee
Taking the invariance of $c$ as an axiom in the theory of
special relativity, one can derive the Lorentz transformation
law without the need for any reference to electric or magnetic 
fields. Maxwell theory on the other hand, as a classical field 
theory, regards the dielectric constant $\epsilon_0$ and the
magnetic permeability $\mu_0$ as independent properties of the 
physical vacuum. Light waves are solutions of the Maxwell theory, 
and the speed of light is calculable. The proposed relationship 
between quantum field theory and an underlying classical field 
theory in a higher dimension is analogous.

In this letter we report the results of numerical simulations 
of a U(1) lattice gauge theory both in 4 and 4+1 dimensions. 
The 4-dimensional Euclidean quantum gauge theory was simulated 
by standard quantum Monte-Carlo techniques. The 4+1-dimensional 
classical theory was simulated by numerical solution of the
differential equations describing its Hamiltonian dynamics. 
In both cases, $4^4$ hypercubic lattices were employed. Lattice 
links beginning at point $x$ and pointing in the $\mu$ direction 
are associated with phases, $A_{\mu}(x)$, of unimodular complex
numbers $U_{x,\mu}=\exp[igaA_{\mu}(x)] \in {\rm U}(1)$
in these theories.

\section{Classical and Quantum Lattice Gauge Models}

In order to discuss the lattice regularization of classical and 
quantum field theory in parallel, we use a common notation as far 
as possible. Our starting point is the definition of the link 
variables $U$, which are always elements of the local U(1) group, 
but the interpretation, and hence the decomposition, of the phase 
can be different. In order to relate a physical interpretation to 
the lattice model phase, $gaA$, we consider the product of charge 
and vector potential, $qA$ (in the CGS system with \hbox{$c=1$)}, 
which is the interaction energy associated with the field living 
on a link of length $a$ with a physical test charge $q$. This
expression must be divided by another quantity with the dimension 
of an energy to generate a dimensionless phase variable. This 
quantity, the unit link energy $\epsilon$, is used as an energy 
``standard'', but it does not appear in the equations of motion
of the theory. Also, we are free to choose $\epsilon$ differently 
in the quantum and in the classical theory.  The general link 
variable is now given by
\be
 U_L = \exp( i qA/\epsilon).
\label{GENER-LINK}
\ee
In the standard quantum lattice gauge field theory the energy 
scale associated with a link of length $a$ is
\be
\epsilon_{\rm Q} = \hbar / a .
\label{MC-EN}
\ee
In a classical theory, however, no reference to the Planck constant 
$\hbar$ is allowed. In this case we will use instead the Coulomb
energy associated with two charges $q_{\rm K}$ at the endpoints of 
the link of length $a$, which in three spatial dimensions is given by
\be
\epsilon_{\rm K} = q_{\rm K}^2/a.
\label{EN-KL}
\ee

By comparing the classical ($K$) and quantum ($Q$) systems
of U(1) lattice gauge fields, we shall consider configurations 
characterized by the same link phase $gaA$. Using a common 
lattice spacing $a$ and a common coupling strength $g$, this 
convention naturally leads to the consideration of the same 
vector potential field, $A$, in the continuum limit. We can
express this correspondence by the relation
\be
q_{\rm Q} / \epsilon_{\rm Q} = q_{\rm K}/\epsilon_{\rm K} = ga.
\label{FIX-ga}
\ee
As a consequence the classical charges as reference sources 
for the unit link energy are interpreted differently in the 
two cases: $q_{\rm Q}=g\hbar$ in the quantum theory and 
$q_{\rm K}=1/g$ in the classical lattice theory. In a sense, 
the two approaches are dual to each other, with 
$q_{\rm Q} q_{\rm K} = \hbar$, i.e. a high value of the 
charge corresponds to strong coupling in the quantum, but to 
weak coupling in the classical theory. 

We note that the conventions chosen by us here are not unique. 
Alternative conventions, e.g. relating $\epsilon_{\rm K}$ to 
a classical charge in four spatial dimensions, as 
$\epsilon_{\rm K}=q_d^2/a^2$ are also possible, but they would
not lead to different conclusions. In the end, the classical
theory as well as the quantum theory are defined through the
interactions of the dimensionless link variables $U$. The
classical Hamiltonian $H_{\rm K}[U]$, being scale invariant, 
is a nonlinear function of the link variables, multiplied by
a constant parameter carrying the dimension of energy. 
Likewise, the action of the quantum theory $S_{\rm Q}[U]$,
which depends on the scale only through the gauge coupling
$1/g^2(a)$ as an overall factor, has the dimension of an
action. If, as we will show below, the two theories are
related to each other, this implies a relationship between
the dimensionful factors of $H_{\rm K}$ and $S_{\rm Q}$, 
which does not depend on the conventions used in the definition 
of the phase of $U$.

Physical quantities are related to the oriented product of link
variables around an elementary square, a plaquette. Using the
notion of lattice forward derivatives,
$a\partial_{\mu}f = f(x+ae_{\mu}) - f(x)$, 
(where $e_{\mu }$ is a unit vector in the corresponding direction)
the plaquette phase sums are associated with the local rotations
of the vector potential, with the field strength tensor
\be
F_{\mu\nu}(x) = \partial_{\mu}A_{\nu}(x) - \partial_{\nu}A_{\mu}(x).
\ee
An elementary plaquette variable is related to a component
of the field strength  tensor:
\be
U_{x,\mu\nu} = \exp\left( iga^2 F_{\mu\nu}(x)\right),
\ee
and the lattice sum over all plaquettes
\be
\Sigma_{\rm P} = \sum_x \sum_{\mu > \nu} 
  \left(1 - {\rm Re} \: {\rm tr}(U_{x,\mu\nu})\right)
\ee
is related to the physical energy or action, depending on the 
dimensionality of the considered space.  For the U(1) group,
\be
1 - {\rm Re} \: {\rm tr}(U_{x,\mu\nu}) = 
  1 - \cos\left(ga^2 F_{\mu\nu}(x)\right),
\ee
which in the continuum limit $a \rightarrow 0$ approaches
$(1/2)g^2a^4(F_{\mu\nu})^2.$ The plaquette sum approximates
\be
\Sigma_{\rm P} \approx g^2a^4 \sum_{x,\mu\nu} \frac{1}{4}F_{\mu\nu}^2(x).
\ee
Over a 4-dimensional lattice it is proportional to
the action, $\Sigma_{\rm P} = g^2 S_4$, and can be
used to express the logarithmic weight of a configuration:
\be
S_4/\hbar = \frac{1}{g^2\hbar} \: \Sigma_{\rm P} \; 
 	= \; \beta \Sigma_{\rm P},
\ee 
where $\beta$ is a dimensionless constant playing the role
of a (fictitious) temperature.
Note that the fine structure constant can be expressed as
$\alpha = q^2/4\pi\hbar$, yielding the familiar relation
\be
\alpha_{\rm Q} = \frac{g^2\hbar^2}{4\pi\hbar}=
\hbar \frac{g^2}{4\pi} = \frac{1}{4\pi\beta}
\label{Q-alpha}
\ee
for quantum lattice gauge theory, and 
\be
\alpha_{\rm K} = \frac{1}{4\pi\hbar g^2} = \frac{\beta}{4\pi},
\label{K-alpha}
\ee
for the classical theory. The duality of these interpretations
is manifest here, as well.

In the (4+1)-dimensional classical lattice gauge theory the 
summation over all pla\-quettes of a 4-dimensional spatial lattice
is proportional to the (four-dimensional) magnetic energy. Since 
the classical interpretation cannot refer to the Planck constant, 
we arrive at the following expression for the energy by using 
$\epsilon_{\rm K}$ as the energy unit:
\be
E_5^{\rm magn} = \epsilon_{\rm K} \Sigma_{\rm P} =
	\frac{1}{g^2a} \Sigma_{\rm P}.
\label{MAG-EN}
\ee
Here we used (\ref{FIX-ga}) to eliminate $q_{\rm K}$ from the
definition (\ref{EN-KL}) of $\epsilon_{\rm K}$. In the continuum 
limit, the magnetic energy becomes
\be
E_5^{\rm magn} \approx a^4 \sum_{x,\mu\nu} \frac{1}{4a} F^2,
\ee
confirming that the physical gauge fields in 4+1 dimensions,
$A^{(4+1)}$ are related to those in 4 Euclidean dimensions,
$A^{(4)}$, by a factor $a^{-1/2}$. 

We now come to our main point. The conjecture $\hbar = aT$ is 
supported if the quantum and classical lattice simulations can 
be brought into a correspondence relating configuration weights,
and thereby all physical expectation values, by
\be
\exp(-S_4/\hbar) \propto \exp(-E_5^{{\rm magn}}/T) 
		 \propto \exp(-\beta \Sigma_{\rm P}).
\label{E-S}
\ee
We have used proportionality signs in this relation, because
the overall normalization of the weights is not relevant.
Also, the relation (\ref{E-S}) defines the temperature $T$
as the equipartitioning energy for the magnetic energy of
the classical lattice field. This designation obviously
requires that the classical theory is ergodic, so that a
single field trajectory generates configurations with a
microcanonical equilibrium distribution.

Quantum Monte Carlo algorithms generate configurations of 
lattice link elements according to the Gibbs weight 
$w[U] \propto e^{-\beta\Sigma_{\rm P}} = e^{-S[U]/\hbar}$.  
The classical Hamiltonian approach, on the other hand, 
divides the Hamiltonian into its electric and magnetic parts:
\be
H =  a^4 \sum_{x,\mu} \frac{1}{2a} \dot{A}^2 + E^{\rm magn}[U]. 
\label{H-class}
\ee
This formula leads to the following dimensionless Hamiltonian
lattice model, where the time is measured in lattice spacing
units $a$ and time derivatives take the dimensionless form 
$a(\partial /\partial t)$:
\be
g^2aH = \frac{1}{2} \sum_{\rm L} {\rm tr}
     \left(a\dot{U}a\dot{U}^{\dag}\right) + \Sigma_{\rm P}.
\label{SCALING-H}
\ee

Now the link variables $U_{x,\mu}$ are still defined on a 
four-dimensional lattice, but they are functions of an additional,
continuous and scaled time variable $t/a$. The evolution of the 
lattice configuration occurs in this fifth dimension, usually 
called ``fictitious'' time in the context of stochastic quantization. 
In the present context, however, the time-like dimension is 
considered as physical, albeit unobservable at long distances and 
low frequencies. One reflection of this difference is that we do 
not add an external heat bath or white noise to the classical 
dynamical equations; we here consider pure classical Hamiltonian 
dynamics.

\section{Technical Aspects of the Simulations}

We now explain some technical aspects of our simulations, before
we report the results. For the purpose of generating generic
initial field configurations, and as a matter of convenience, we
prepared the initial configuration for the classical Hamiltonian 
simulation by Monte-Carlo ``heating'' on a five-dimensional lattice. 
The $\beta_5$ parameter chosen for this simulation determines the 
average value of the action $S_5$ and of the five-dimensional 
electric and magnetic energies. The selected configuration was 
then converted into the initial data for a four-dimensional, real-time 
lattice calculation as follows: The space-space-like plaquettes
and links were taken without change from the four-dimensional
hypercube located at $x_5=0$. The periodic boundary condition 
in the $x_5$-direction allowed us to identify the link phase for 
$x_5/a_5=-1$ with those at $x_5/a_5=N_5-1$ for the construction
of the initial values for the time derivatives $\dot{U}$ on the 
hypercube in terms of the link variables attached to the plane
$x_5=0$. This construction also used the link phases in the $x_5$ 
direction to make the time derivatives gauge covariant. By 
construction, the $\dot{U}$ variables then are orthogonal to the
corresponding $U$ variables, as required by the Hamiltonian time 
evolution. We now present the details of our algorithm for the
construction of the initial gauge field configurations.

We denote by $U_+$ the triple product of link variables including the 
value at $x_5/a_5=1$ (the first argument refers to the site coordinates 
$x=\{x_1,\cdots,x_4\}$; the second argument indicates the value of
$x_5/a_5$):
\be 
U_+ = U_5(x,0) U_x(x,1) U^{\dag}_5(x+a_s,0) 
\ee
and similarly by $U_-$ that product including the previous-time value
\be 
U_- = U^{\dag}_5(x,-1) U_x(x,-1) U_5(x+a_s,-1) 
\ee
Both triples close a plaquette with the $U=U_x(x,0)$ link. These 
plaquettes encode the electric field values at $x_5 \pm a_5/2$.
Using the scalar product notation for U(1) group elements:
\be
\langle A,B\rangle = \cos( \phi_A-\phi_B ),
\ee
we can determine the electric field energy densities
\hbox{$\varepsilon_{\pm}= \frac{1}{2}E_x^2(x,\pm\frac{a_5}{2})$}
as follows:
\ba 
\langle U_+, U \rangle & = & 1 - (a_5a_s)^2 \varepsilon_+,
\NL
\langle U_-, U \rangle & = & 1 - (a_5a_s)^2 \varepsilon_-. 
\ea
We also define an interpolating electric energy density $\varepsilon$
by using a double-sized plaquette:
\be
\langle U_+, U_- \rangle  = 1 - (2a_5a_s)^2 \varepsilon .
\ee

The momentum $P$ stands for $\dot{U}dt$ in the classical calculation.
We relate $\dot{U}$ to the link variables of the five-dimensional 
lattice by identifying $dt=a_5$ for the initial state of the
Hamiltonian simulation. Later on we use this initial value of
$P$ and $U$ for updates in much smaller steps than the original
lattice spacing, $dt/a_5 \ll 1$. (Typical values are $0.01$ and 
$0.001$.) The initial $P$ can be expressed as a linear combination 
of the $U$ and $U_{\pm}$ link variables; to leading order it is a 
difference (with a possible admixture of second derivative),
\be
P  = \frac{1}{2}(U_+ - U_-)  + \frac{\delta}{2} ( U_+ + U_- - 2U).
\ee
The parameter $\delta$ is obtained from the orthogonality constraint
\be 
\langle P, U \rangle = \frac{1}{2} (a_5a_{\rm s})^2 
  (-\varepsilon_+ + \varepsilon_- ) - \frac{\delta}{2} 
  (a_5a_{\rm s})^2 (\varepsilon_+ + \varepsilon_- ) = 0. 
\ee
This equation can be easily solved for $\delta$, and $P$ is obtained as
\be 
P = \frac{ \varepsilon_- U_+ - \varepsilon_+ U_- 
          - (\varepsilon_- - \varepsilon_+ )U }
         {\varepsilon_- + \varepsilon_+} .
\ee
The electric energy per link is related to the dimensionless kinetic term
\be
 \frac{1}{2} \langle P, P \rangle \, = \, a_5^2a_{\rm s}^2 
   \frac{4\varepsilon_- \varepsilon_+ \varepsilon}
        {(\varepsilon_- + \varepsilon_+)^2} .
\ee

The lattice simulation programs were coded in $C++$ constructing
classes for arbitrary sized and arbitrary dimensional lattices
with an established site, link, plaquette and elementary cube
index system. In particular indices for calculating gradient,
rotation and divergence were taken care of. Classes for
group variables were also constructed with group multiplication,
Haar measure and random generation functions. In case of U(1)
a single real phase represented the group variable. Data 
i/o handling was coded in compressed binary form with a standardized
header of lattice size information.

Algorithms for cold, hot and $\exp(-\beta(1-\cos\varphi))$
distributed initialization, as well as  Monte Carlo heating
algorithms based on the original Metropolis rejection method
or in another version on a local heat-bath mechanism were implemented
for producing lattice configurations. Functions like the complex
Polyakov line average over the lattice space volume and the action
were included. A conversion program between data on a $d$ dimensional
lattice and a Hamiltonian pendant of the configuration on a $(d-1)$
dimensional lattice (plus an almost continuous time variable)
determined $U$ and $\dot{U}$ values to begin a classical solution
of the equations of motion. These were solved by a program obtaining
updates in a stabilized half-time-step algorithm,
($U \leftarrow U + P/2$, $P \leftarrow P + (dt^2) F$, 
$U \leftarrow U +  P/2$  again), which ensures that energy is
conserved up to a precision of ${\cal O}(dt^3)$. We have used typically
$dt/a=0.01$.

We now give a brief summary of our tests of these programs for performance.
The group multiplication, trace, determinant and Haar-measure routines
were tested on special (unity, $\exp(i\pi)$) and on random phase
elements. The random generators, one using the simple rejection
method with $1$ as majoring function and one using a Gaussian to
decide whether to reject or not, were tested for averages, variances,
auto-correlation and computational time. The transient effects by
Monte Carlo heating were monitored by the average action, by the
Polyakov value squared and by the magnetic monopole density.
An optional acceleration of heating was implemented by the link phase mirroring 
technique inserted between the traditional heat-bath steps.
The usual phase transition as a function of the quantum Monte Carlo
coupling constant has been performed at the expected value for $4^4$,
$6^4$, $8^4$ and $12^3\times 4$ lattices. In the classical equation
of motion solver the energy conservation and the equipartition
of the kinetic and potential energy parts has been monitored.
All averages presented in the following are already screened for transient
effects, which have been removed by 
eliminating early parts of the trajectory.


\section{Numerical results}

\begin{figure}[ht]
\begin{center}
\includegraphics[width=0.45\textwidth]{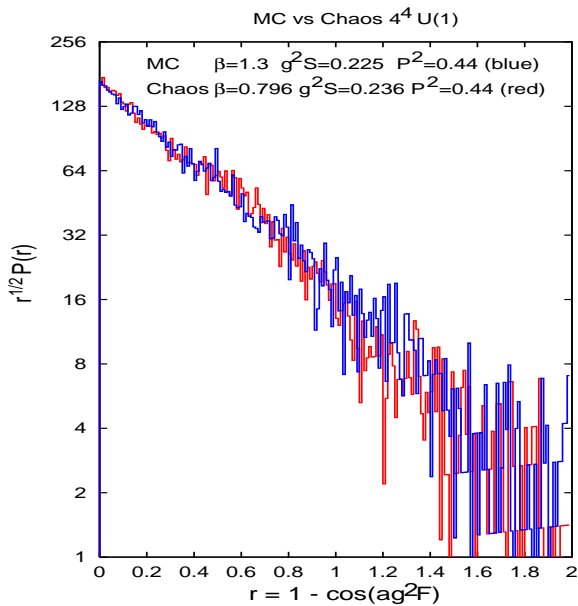}
\qquad 
\caption{ \small The distribution of plaquette values is well
approximated by an inverse square root times an exponential decay law. 
This provides numerical
evidence for the self-thermalization of the classical U(1) lattice
system due to its chaotic dynamics (red histogram), as well as
of the Monte-Carlo heat bath algorithm (blue histogram).
}
\label{EXP}
\end{center}
\end{figure}

Microcanonical equilibration of classical Hamiltonian lattice fields 
has been studied earlier for the non-Abelian groups SU(2) and SU(3) 
in three spatial dimensions \cite{chaos}.
Here we repeat this study for the U(1) system in $4d$ in order to make sure
that this system also approaches a thermal distribution of plaquette
energies during its chaotic evolution. Fig.\ref{EXP} plots the
frequency of occurrence of the $r = 1 - \cos(ag^2F)$ plaquette values
in the allowed interval of $[0,2]$ on a logarithmic scale
for a corner plaquette at different times (sampled at each 5-th
time-step of an evolution of 100.000 steps of each having $dt/a=0.01$).
Due to the periodic boundary conditions there is nothing special
about this position. One observes that the distribution of
plaquette values is well approximated by an exponential law 
multiplied by the inverse square root function (red histogram).
We note by passing that a similar distribution has been observed
in earlier studies of 3-dimensional SU(3) lattice gauge models,
but with a prefactor of $r^3$ instead of $r^{-1/2}$ \cite{stat}. 

For the sake of comparison the same distribution is plotted
for the 4-dimensional quantum Monte-Carlo algorithm at inverse coupling
$\beta=1.3$ (blue histogram). The correspondence is made due to
similar values of the average Polyakov line squared 
($\langle P^2 \rangle_{MC} \approx \langle P^2 \rangle_{\rm Chaos} \approx  0.44$) 
at close values of average plaquette traces ($g^2S_4 \approx 0.225$
and $g^2aE_5^{magn} \approx 0.23$).
The agreement between these two distributions, not only in the shape, but
also in absolute number, already demonstrates that we are producing
instances of lattice configurations with the same probability by applying
either the traditional quantum Monte Carlo or by the chaotic dynamics
method.

\begin{figure}[ht]
\begin{center}
\includegraphics[height=0.45\textwidth,angle=-90]{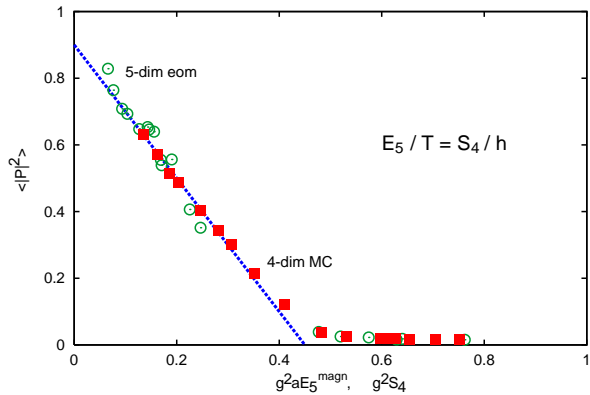}
\caption{ \small The order parameter, the absolute value square of
the Polyakov line averaged over the lattice and over many
configurations obtained by classical Hamiltonian dynamics
is plotted against the magnetic energy of the four-dimensional
spatial lattice (open circles), and the same quantity obtained by traditional 
quantum Monte-Carlo simulation on a Euclidean four-dimensional 
lattice against the four-dimensional action (filled boxes), respectively.
These results coincide, if and only if $E_5 / T = S_4/\hbar$.}
\label{Fig1}
\end{center}
\end{figure}

We now further demonstrate the validity of the relation (\ref{E-S})
by showing the equivalence of the two theories for an important
observable, the expectation value of the modulus squared 
of the Polyakov line on the respective periodic lattices. 
We note that this is the standard order parameter of the 
lattice gauge theory, which vanishes at strong coupling and 
is nonzero below a critical coupling $\beta_c \approx 1$, 
indicating the transition to the Coulomb phase. 
Figure \ref{Fig1} shows the Polyakov line modulus squared
averaged over the 3-volume of the 4-dimensional lattice and
over  20,000 quantum Monte-Carlo configurations (red squares) 
as a function of the average 4-dimensional lattice action per 
plaquette $\Sigma_{\rm P} = g^2 S_4$. 

On the same plot the same order parameter is shown (green dots) 
as a function of the expectation value of the lattice magnetic 
energy $\Sigma_{\rm P} = g^2 a E_{\rm mag}$, per plaquette,
of the classical configuration. The average here was calculated, 
after an additional brief Hamiltonian equilibration on the 
classical four-dimensional lattice, as an ergodic average by 
temporal sampling of a single evolving lattice field configuration. 
We note that these two sets of points follow the same scaling law.

The reason that we do not plot the modulus of the Polyakov line as 
a function of the inverse coupling $\beta$, as it is usually done, 
is that the classical Hamiltonian (\ref{H-class}) does not contain 
such a coupling constant. Its solutions are solely characterized by 
the value of the total energy or, because of its ergodicity properties,
by the average value of the magnetic energy. The fact that our 
simulation points obtained for the quantum field theory on the 
four-dimensional Euclidean lattice and the (4+1)-dimensional classical
theory coincide when we impose the relation (\ref{E-S}) constitutes
the desired evidence. (One might consider relating the two results
via the coupling constant $\beta_5$ used in the generation of the initial
conditions for the classical field configuration. However, this would
be inappropriate, because the five-dimensional lattice provides us 
solely with a convenient algorithm for generating randomized initial 
data. We could have started the classical calculation with an arbitrary 
initial configuration of the same total energy, without any reference
to a five-dimensional lattice, and would have obtained the same
results, albeit after much longer microcanonical equilibration.)

In order to further confirm the validity of the relation between 
the four-dimensional quantum and (4+1)-dimensional classical
lattice U(1) theories it is illustrative to look at scatter plots 
of sampled values of the Polyakov line in the complex plane.
In Figure \ref{Fig2} the results from the four-dimensional 
quantum Monte-Carlo simulation are shown as red dots, and the
results from the (4+1)-dimensional classical Hamiltonian evolution 
for a single trajectory are represented by green dots. 
The correspondence in the four parts of Fig.~\ref{Fig2} is again 
made by selecting pairs of simulations satisfying the relation
(\ref{E-S}). The overlap of the distributions is excellent, both 
for the modulus and the phase of the Polyakov line. Some initial 
points in the center of the rings for supercritical couplings 
$\beta_4 = 1/g_4^2 \ge 1$ are artefacts from an initial heating 
phase in the quantum Monte Carlo simulations. They would be absent, 
if we had started the sampling process at a later time. In the 
classical Hamiltonian evolution the ergodic sampling was started 
also only after a transitory period, in which the electric and 
magnetic field energies equilibrated (in 4+1 dimensions, the ratio 
of electric and magnetic energy is 2:3, not 1:1 as in the usual 
(3+1)-dimensional case).


\section{Conclusion}

In conclusion, we have demonstrated that the mechanism of chaotic 
quantization - conjectured earlier on the basis of non-Abelian 
gauge theories in 3 and 3+1 dimensions - also works for the compact
Abelian lattice gauge theory in 4 and 4+1 dimensions. The 
correspondence between the classical and quantum gauge theories
is given by the general formula $\hbar = a T$, which encodes 
physical properties of the higher dimensional theory into the 
Planck constant. Our result supports the speculation that chaotic
quantization may be a physical mechanism for the quantization of
(gauge) fields. This could make it eventually possible to construct 
a {\it classical} field theory encompassing both, gravity and the 
standard model of particle physics. In this framework, Planck's 
constant $\hbar$ would become a parameter of the low-frequency, 
long-second-time limit of the fundamental classical field theory.

Finally we would like to address the question whether factorizing
the Planck constant would be tantamount to the construction of a 
hidden variable theory. We believe that this is not the case, since 
none of the established rules of quantum physics are violated. The
four-dimensional quantum field theory is given by a Euclidean functional
integral, which is exactly the path integral that defines the vacuum
sector of the compact U(1) gauge theory in (3+1) dimensions. The higher 
dimensional classical dynamics ``acts'' as a quantum field theory in 
four Euclidean dimensions. 

{\it Acknowledgments:} This work was supported in part by DOE
grant FG02-96ER-40945 and the Hungarian National Research Fund
OTKA with the contract number T 034269.


\appendix
\section{Chaotic quantization of harmonic oscillator}

In order to help to gain a deeper insight into the chaotic quantization
mechanism we present here a limiting case, the harmonic oscillator.
Of course, pure lattice gauge theories cannot be used as models of the
oscillator because they lead to massless particles in the continuum
limit with a trivial dispersion relation $\omega=0$ in the
$0+1$ dimensional case of ordinary quantum mechanics.
An additional ingredient is needed for this purpose; we choose a complex
scalar (Higgs) field which generates mass for the U(1) fields leading
to the desired harmonic oscillator description in the low dimensional
case. We consider the following action,
\be
 S = \int \left( - \frac{1}{4} F_{\mu\nu}F^{\mu\nu} + 
 	\frac{1}{2} |D_{\mu}\Phi|^2    - V(|\Phi|)\right) d^4x
\ee
where the Higgs potential $V$ prefers a non-trivial vacuum state
$|\Phi(t,x)|=R$. The Euclidean lattice version of this theory
is investigated on a narrow, $N\times 1$ stripe in the $t-x$ plane
(here $t$ denotes the ''ordinary'' Euclidean time direction), 
having a periodic boundary condition in the dummy
$x$ direction. The link phases starting at sites $t_j=j a$ and pointing
in the (extremely short) $x$ direction are denoted by $\varphi_j$,
the ones pointing in the $t$ direction by $p_j$. The complex fields
at site positions are given by $\Phi(t_j)=R_je^{i\vartheta_j}$.
The lattice action is decomposed as
\be
 S_{E,lat} \: = \: S_g + S_x + S_t,
\ee
with
\ba
S_g &=& \frac{1}{g^2} \sum_j 1 - \cos(\varphi_{j+1}-\varphi_j), \NL
S_x &=& \frac{a^2}{2} \sum_j R_j^2 \left|e^{i(\varphi_j+\vartheta_j)}-e^{i\vartheta_j}  \right|^2, \NL
S_t &=& a^2\sum_j \left|R_{j+1}e^{i(\vartheta_{j+1}+p_j)}-R_je^{i\vartheta_j} \right|^2.
\ea
In the Higgs phase with a Lorentz-invariant (in the case of $0+1$ dimensional
quantum mechanics, just simply static) modulus $R_j=R$, terms containing
$\varphi_j$ decouple from the rest. This constitutes the lattice action
for the harmonic oscillator, reducing the dynamics to that of a long chain
of minimal plaquettes in the $t$ direction:
\be
S_{lat}[\varphi] = \frac{1}{g^2} \sum_j(1-\cos(\varphi_{j+1}-\varphi_j)) +
  		(aR)^2\sum_j(1-\cos\varphi_j).
\label{LAT-OSC}
\ee
With the correspondence {$\omega=gR$} and 
{$\varphi_j = g q_j \, \sqrt{m/a}$} we get 
the familiar action of the harmonic oscillator in the small phase 
(either weak coupling, or small amplitude, or small oscillator mass) limit
\be
 S_{lat}[q]  =  a \sum_j\left(\frac{m}{2}\left(\frac{q_{j+1}-q_j}{a}\right)^2
 		+ \frac{m\omega^2}{2} q_j^2 \right).
\ee
For $a \rightarrow 0$ this leads to the continuum Euclidean action of a single 
harmonic oscillator:
\be
S_c[q] \: = \: \int \left(\frac{m}{2}\dot{q}^2 + \frac{m\omega^2}{2}q^2 \right) dt.
\ee
We now turn to the analysis of the Hamiltonian description of this lattice
chain model (\ref{LAT-OSC}) describing its classical evolution in 
a second time $s$.  We use the Hamiltonian
\be
 H_5 = \frac{a}{2g^2} \sum_j \left(\frac{d\varphi_j}{ds} \right)^2
 	+ \frac{1}{a} S_{lat}[\varphi].
\ee
The classical equation of motion, presented in a scaled time variable $s/a$
becomes
\be
\frac{d^2\varphi_j}{d(s/a)^2} = \sin(\varphi_{j+1}-\varphi_j) - 
 \sin(\varphi_j-\varphi_{j-1})-(a\omega)^2\sin\varphi_j.
\ee
The only parameter in this $s$-dynamics is the scaled frequency
$G=a\omega$. In a molecular field approximation each link
phase (or plaquette phase) would behave like a pendulum with a
harmonically perturbed pivot point. Such systems are
known to be chaotic \cite{PEND-TO-CHAOS}.

In order to understand how the classical chaotic mechanics leads to
a distribution in the second time $s$, which resembles that
distribution applied in Euclidean quantum field theory, the key
property is the relation between the equations of motion and a
Fokker-Planck-Kolmogorov equation. This relation has also been
studied at length in chaotic dynamics and the circumstances when
the chaotic motion can be replaced by a diffusion in phase space
governed by a Fokker-Planck type equation have been investigated.
Here we follow a presentation given in Chapter 6 of Ref.\cite{PEND-TO-CHAOS}.

First the differential equations are solved in discrete time steps,
$\Delta s$, generating a mapping in the phase space:
\ba
\overline{q_i} &=& q_j + \frac{p_j}{m} \Delta s, \NL
\overline{p_i} &=& p_j + F_j \Delta s
\ea
with the following force in the lattice chain model
\be
 F_j = - \frac{g}{a} \sqrt{\frac{m}{a}} \frac{\partial}{\partial \varphi_j}
    S_{lat}.
\ee
The mapping should be in action-angle variables; the lattice model
is readily formulated in terms of the angles $\varphi_j$. The corresponding
scaled action variable we denote by $I_j= a (d\varphi_j/ds)$.
The (2-time) Hamiltonian contains the following kinetic energy term
\be
 H_5^{kin} = \frac{m}{2} \sum_i \left(\frac{dq_i}{ds}\right)^2 =
 \frac{1}{g^2a} \frac{1}{2} \sum_i I_i^2.
\ee
allowing us to write the mapping in dimensionless action -- angle
variables as follows
\ba
\overline{\varphi_j} &=& \varphi_j + \frac{\Delta s}{a} I_j, \NL
\overline{I_j} &=& I_j - \frac{\Delta s}{a} g^2 
	\frac{\partial}{\partial\varphi_j} S_{lat} .
\ea
An equivalent Fokker-Planck-Kolmogorov equation is then valid 
for the distribution of $I_i$ in a long-term sampling of  phase
space trajectory:
\be
 \frac{\partial {\cal F}}{\partial s} = -
 \frac{\partial}{\partial I_i} (A_i {\cal F}) + \frac{1}{2}
 \frac{\partial}{\partial I_i} \frac{\partial}{\partial I_j} 
 (B_{ij}{\cal F})
\ee
with the following drift and diffusion coefficients
\ba
A_i &=& \frac{a}{\Delta s} \langle \langle \overline{I}_i - I_i \rangle\rangle, \NL
B_{ij} &=& \frac{a}{\Delta s} \langle \langle (\overline{I}_i - I_i) 
 (\overline{I}_j - I_j) \rangle\rangle
\ea
Here the averaging over different observation moments $s$ and different
trajectories (wherever started) can be replaced by an average
over the angle variables. This is a central statement of chaotic dynamics,
giving a statistical physics perspective to deterministic dynamical systems.
This is as well the key mechanism beyond the chaotic quantization,
arriving at a Fokker-Planck correspondence, which used to be the starting point
of the stochastic quantization method.

For our lattice oscillator model we get $A_i=0$ and
\be
B_{ij}=\frac{\Delta s}{a}\left((1+\frac{(a\omega)^2}{2})\delta_{ij}-\frac{1}{2}
	(\delta_{i-1,j}+\delta_{i+1,j})\right).
\label{DIFFU}
\ee
meaning that the diffusion matrix, the coefficient of the scaled time-step
$\Delta s/a$ in the matrix $B_{ij}$ is readily recognized as the 
the harmonic oscillator matrix in the quantum mechanical 
Euclidean path integral formalism.

At this point we still seek to determine the long second-time limit
of the diffusion matrix $B_{ij}$. This can be understood with the help
of the Green-Kubo formula \cite{Dorfman} which connects the microscopical
dynamics with the linear response approximation. The diffusion-like
Fokker-Planck equation is actually the leading order cumulant
expansion to the more general case, the Fourier transform of the
probability is taken in the infrared limit associated to the
basic action variables:
\be
 {\cal F}_k = \left\langle e^{ik_j\Delta I_j} \right\rangle \approx
 \exp \left( -k_i \langle \Delta I_i \Delta I_j \rangle k_j \right).
\ee
Back Fourier transformation leads then to the second derivative
according to the $I_j$ variables; the first derivative (drift) term
vanishes for time reversal dynamics, such as governed by conservative
Hamiltonians. The diffusion coefficient in scaled time $s/a$ becomes
\be
B_{ij} = \left\langle a \frac{\Delta I_i}{\Delta s} \Delta I_j \right\rangle
   =  \int_0^{s/a} \langle \dot{I}_i(s)\dot{I}_j(s') \rangle ds'. 
\ee
Assuming now that the correlation of the action variables depends only
on the absolut value of second-time difference, as it is the case both
in stochastic and chaotic processes, we arrive at
\be
B_{ij} = 
     \int_0^{s/a} \langle \dot{I}_i(s')\dot{I}_j(0) \rangle ds'. 
\label{BINT}
\ee
The characteristic correlation of the force ($\dot{I}$) is an exponential
decay both in the stochastic, as well as in the chaotic quantization.
In the former case a Langevin equation with white noise leads to
forgetting, in the latter case the instability of nearby trajectories
expressed by positive Lyapunov exponents. The general behavior is then
\be
 \langle \dot{I}_i(s')\dot{I}_j(0) \rangle = e^{-\gamma s'}
 \langle \dot{I}_i(0)\dot{I}_j(0) \rangle.
\ee
The second-time integral in (\ref{BINT}) can now be analytically done
leaving us with
\be
B_{ij} = \frac{1-e^{\gamma s}}{a\gamma}
\ee
For a short s-time evolution it gives back the result of (\ref{DIFFU}),
in the long term, however, leads to a constant times the matrix
describing the oscillator's path integral in ordinary quantum mechanics:
\be
 \lim_{s \rightarrow \infty }B_{ij}=
 \frac{1}{a\gamma}\left((1+\frac{(a\omega)^2}{2})\delta_{ij}-\frac{1}{2}
	(\delta_{i-1,j}+\delta_{i+1,j})\right).
\label{FINAL-DIFFU}
\ee
This proves our conjecture for
the harmonic oscillator, without extended numerical simulations.

Finally some remarks are in order to explain how dimensionful scales
and parameters enter in the interpretation of lattice model results.
Of course, both the familiar quantum Monte Carlo and the second-time
Hamiltonian equation of motion (shortly EOM) method deals with scaled
quantities on the lattice (and with a scaled time) not having any
length (fm) or energy (MeV) dimensions. Besides the lattice spacing  
$a$, giving the unit of length, either the unit of quantum action,
$\hbar \approx 197$ MeVfm is given, or in the EOM method instead of
$\hbar$ an energy, which is conserved in total, is given.

The study of the diffusion coefficient in phase space describes the
chaotic evolution in mid terms, the elapsed second time $\Delta s$
to be taken between an autocorrelation time $s_c$ when initial phase points
are remembered by the evolution trajectory (there is an exponential
forgetting of nearbyness of the initial conditions set by the leading
Lyapunov exponent) and the longer time when equipartition completes.
As in the case of simple Brownian motion, the kinetic energy
starts to grow linearly for short time $s$ but levels off at an
equipartition value, in our case we have
\be
\langle \frac{1}{2g^2a} \sum_i I_i^2 \rangle
= N \frac{T}{2} ( 1 - e^{-\gamma s} )
\rightarrow N T / 2,
\ee
where $N$ is the ordinary time extension of the path integral lattice.
(This is the correct scaling of Brownian motion parameters with "volume".)
Expressing the product of the equipartition temperature $T$
and the inverse  equipartition time $\gamma$ in a lattice scaled
fashion we arrive at
\be
 aT = \frac{1}{g^2N} \sum_i B_{ii}.
\ee
containing the average eigenvalue of the diffusion matrix. 
On the other hand the long time equipartition reaches
\ba
\langle \frac{1}{2g^2a} \sum_i I_i^2 \rangle \rightarrow \frac{1}{2} T N, \NL
\langle \frac{1}{a} S_{lat} \rangle \rightarrow \frac{1}{2} T N,
\ea
while the distribution ${\cal F}$ approaches the canonical one
\be
{\cal F} \propto e^{-N} = e^{-\langle H_5 \rangle /T} = 
    e^{-2\langle S_{lat}/(aT) \rangle}
\ee
This canonical distribution is eventually interpreted as the Euclidean
quantum wave function, ${\cal F} \sim \exp(-S/\hbar)$,
(continuing back to real time it would be $\exp(iS/\hbar)$
constituting the solution of the Schr\"odinger equation).
This interpretation of the chaotic distribution allows us to connect
the Planck constant experienced in the 1-time quantum mechanics
with parameters of the phase space diffusion of the 2-time Hamiltonian:
\be
 \hbar = \frac{1}{2} aT = \frac{1}{2g^2} \frac{1}{a\gamma} 
	\overline{\lambda}(B).  
\ee
For the oscillator in the $N \rightarrow \infty$ limit the average eigenvalue
becomes $\overline{\lambda}(B)=(2+(a\omega)^2)$ leading to
\be
 \hbar = \frac{1}{g^2} \frac{1}{a\gamma} \left( 1 + \frac{(a\omega)^2}{2} \right)
\ee
or with $a \rightarrow 0$ at finite $g$ and $\omega=gR$ values, the same
value as for free fields,
\be
 \hbar =  \frac{1}{g^2} \, \frac{1}{\gamma a}.  
\ee
relating Planck's constant with the leading Lyapunov exponent of the
chaotic dynamics in s-time.

For the classical simulation input is the scaled energy,
$g^2aE_5$ which is conserved, and the lattice size $N$.
For the quantum MC simulation input is $\beta=1/(g^2\hbar)$
and the size $N$.
From the above equipartition by chaotic dynamics we got
$g^2\hbar = 1/(a\gamma) = g^2aE_5/(2N)  $.
Since by construction the half of $E_5$ is $S_4/a$ at equipartition,
we tend to a final state where {$\langle S_{lat}\rangle \: = \: N \hbar  $ } .
As we equipartition energy in $5 d$, making a temperature $T$,
with proper scaling we create an action in $4 d$
with an $\hbar$ value per degree of freedom.
In this sense the  value of $\hbar$ is determined by the scaled total 
energy per degree
of freedom into the classical EOM simulation. Its interpretation
and descendence are, however, quite untraditional in chaotic
quantization.


\begin{widetext}
\begin{figure}[ht]
\mbox{\includegraphics[height=0.45\textwidth,angle=-90]{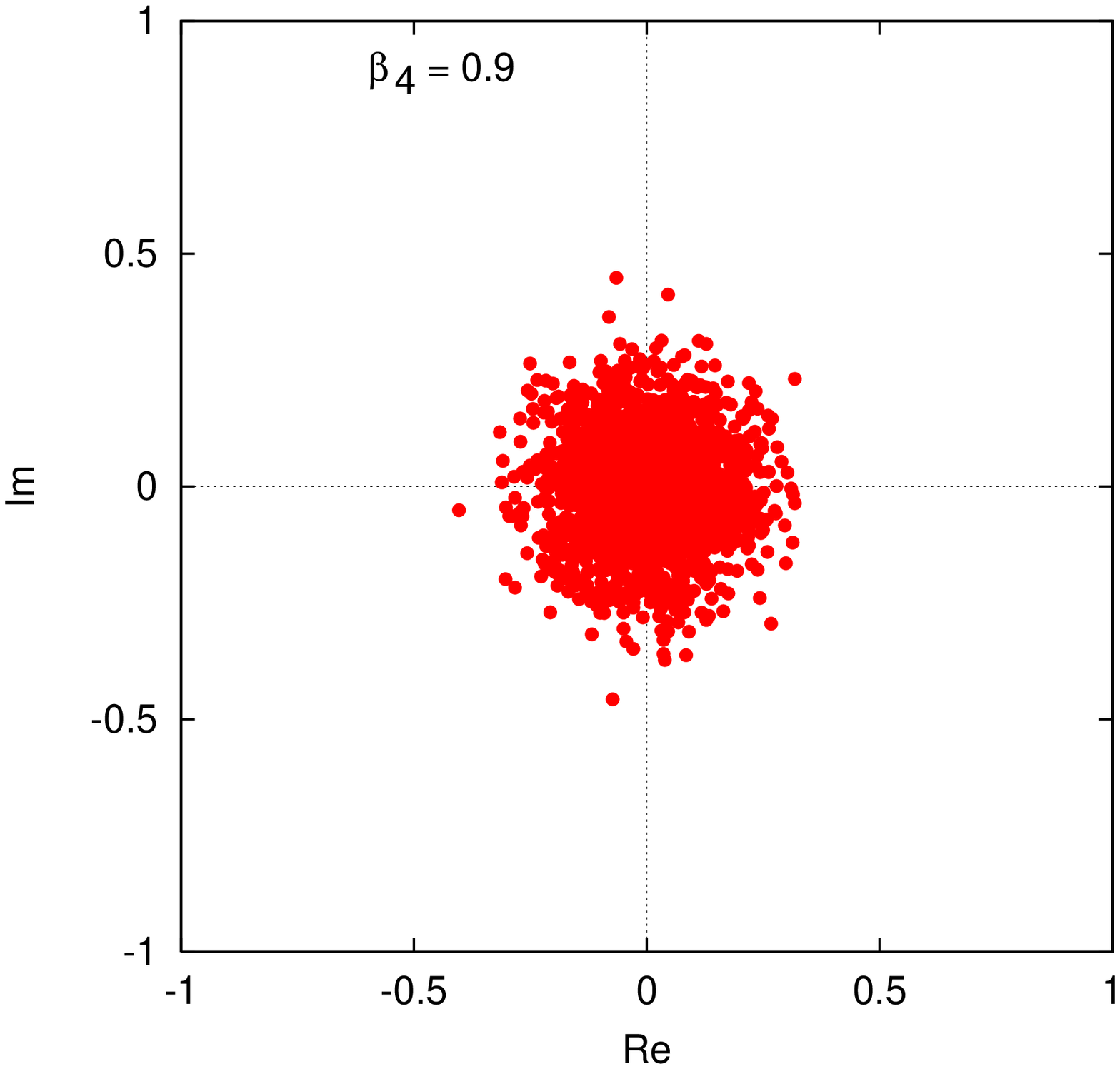}
\includegraphics[height=0.45\textwidth,angle=-90]{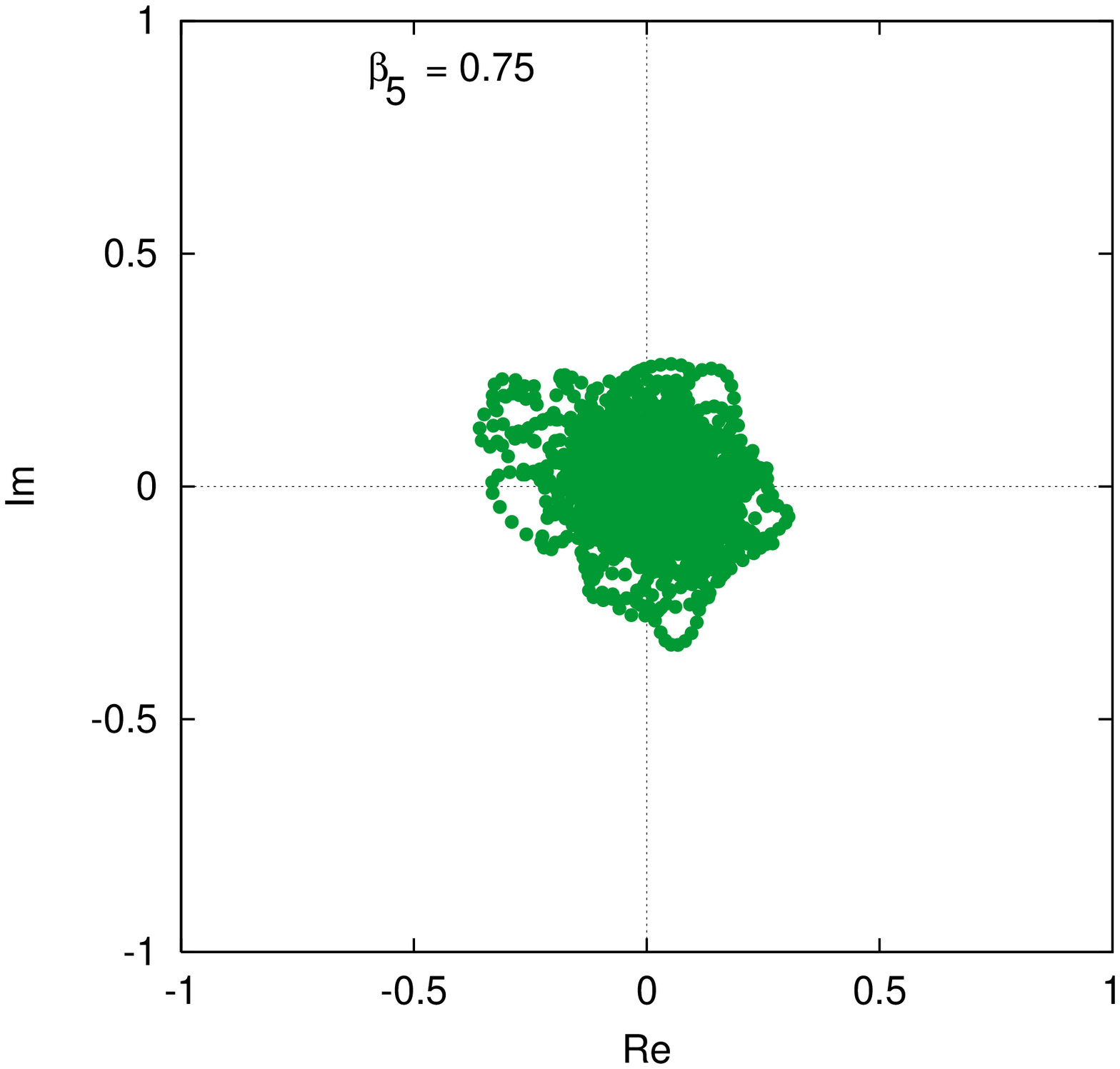}}
\mbox{\includegraphics[height=0.45\textwidth,angle=-90]{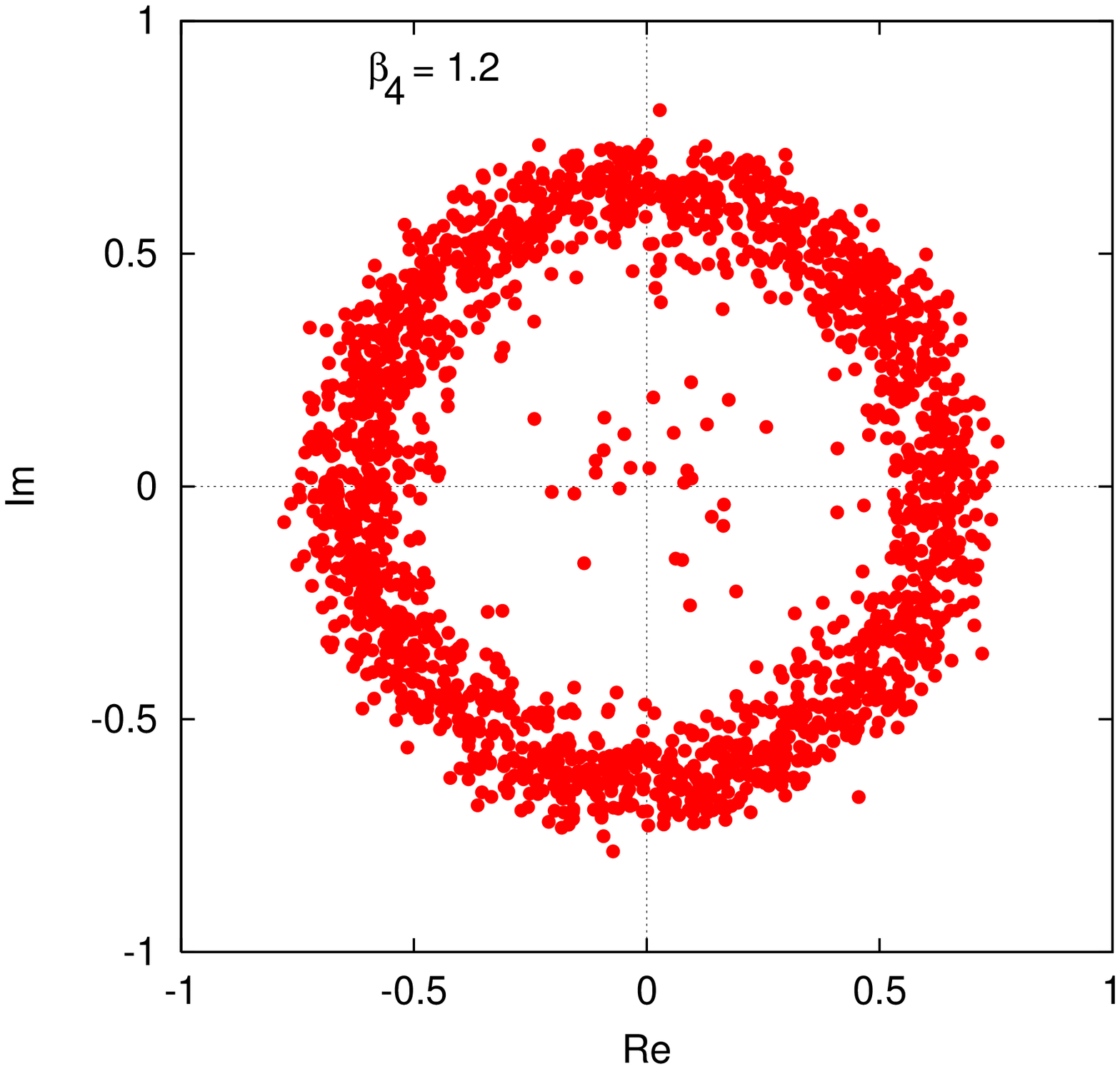}
\includegraphics[height=0.45\textwidth,angle=-90]{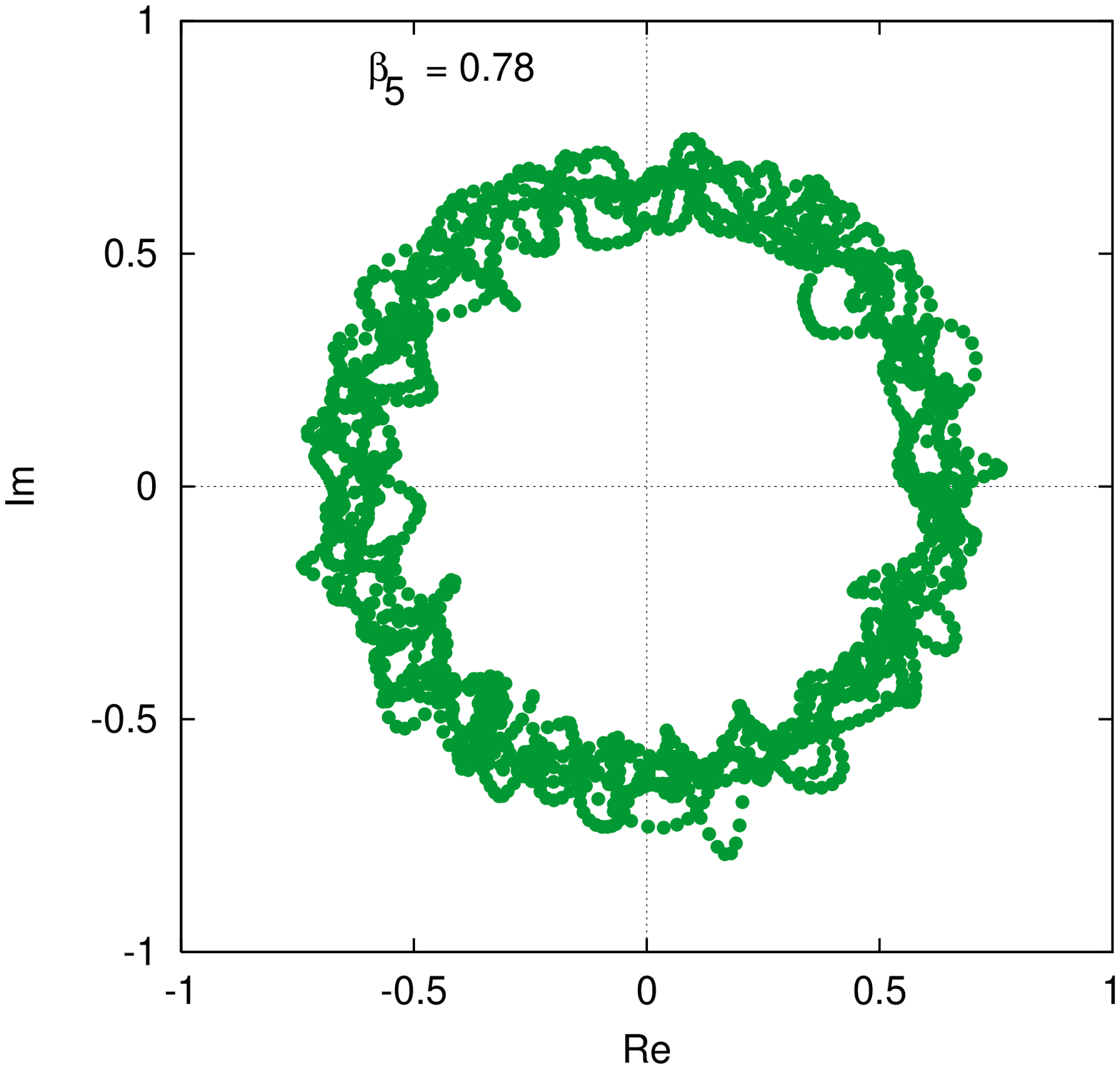}}
\caption{
Complex Polyakov line values from 4-dimensional quantum Monte Carlo 
simulation (red dots) and from (4+1)-dimensional classical Hamiltonian 
equation of motion (green dots) at $aE_4^{mag}/T = S_4/\hbar$.}
\label{Fig2}
\end{figure}
\end{widetext}

\begin{thebibliography}{99}

\bibitem{MST81}
S.~G.~Matinyan, G.~K.~Savvidy, N.~G.~Ter-Arutyunyan-Savvidy:
Sov.\ Phys.\ JETP {\bf 53}, 421 (1981).

\bibitem{BMM95}
T.~S.~Bir\'o, S.~G.~Matinyan, B.~M\"uller,
{\em Chaos and Gauge Field Theory} 
(World Scientific, Singapore 1994).

\bibitem{chaos}
B.~M\"uller, A.~Trayanov, 
Phys.\ Rev.\ Lett.\ {\bf 68}, 3387 (1992);
C.~Gong, Phys.\ Rev.\ D {\bf 49}, 2642 (1994);
J.~Bolte, B.~M\"uller, A.~Sch\"afer, 
Phys.\ Rev.\ D {\bf 61}, 054506 (2000).

\bibitem{BMM01}
T.~S.~Bir\'o, S.~G.~Matinyan, B.~M\"uller,
Found.\ Phys.\ Lett.\ {\bf 14}, 471 (2001);
arXiv:hep-th/9908031, arXiv:hep-th/0301131.

\bibitem{Beck95}
C.~Beck, Nonlinearity {\bf 8}, 423 (1995).

\bibitem{Dorfman}
see e.\ g.\ J.~R.~Dorfman,
{\it An Introduction to Chaos in Nonequilibrium Statistical Mechanics},
Cambridge Lecture Notes in Physics {\bf 14}
(Cambridge University Press, Cambridge, 1999).

\bibitem{tHooft}
G.~'t~Hooft: Class.\ Quant.\ Grav.\ {\bf 16}, 3283 (1999);
arXiv:hep-th/0003004, arXiv:hep-th/0104219, arXiv:hep-th/0104080.

\bibitem{PW81}
G.~Parisi, Y.~S.~Wu, Sci.\ Sin.\ {\bf 24}, 483 (1981).

\bibitem{U1}
H.~Markum, R.~Pullirsch, W.~Sakuler, 
arXiv:hep-lat/0205003, arXiv:hep-lat/0209039;
B.~M.~Gripaios, Phys.\ Rev.\ D {\bf 67}, 025023 (2003);
T.~S.~Bir\'o, H.~Markum, R.~Pullirsch, W.~Sakuler,
arXiv:hep-lat/0210020.

\bibitem{stat}
T.~S.~Bir\'o, C.~Gong, B.~M\"uller, A.~Trayanov, 
Int.\ J.\ Mod.\ Phys.\ C {\bf 5}, 113 (1994).

\bibitem{PEND-TO-CHAOS}
R.~Z.~Sagdeev, D.~A.~Usikov, G.~M.~Zaslavsky,
{\em Nonlinear Physics: From the Pendulum to Turbulence and Chaos},
(Harwood Academic Publishers, Chur, Switzerland, 1988).

\bibitem{XY-MODEL}
S.~G.~Matinyan, Y.~J.~Ng, J.Phys.A: Math.Gen. {\bf 36}, L417, 2003
and references therein.



\end{thebibliography}
\end{document}